\documentclass[aps,prx,twocolumn,a4paper,superscriptaddress]{revtex4-2}
\pdfoutput=1
\usepackage{float}
\usepackage{xspace}
\usepackage{bm}

\usepackage{color}
\usepackage[latin1]{inputenc}
\usepackage{graphicx}
\usepackage{amsmath,amssymb}
\usepackage{booktabs}
\usepackage{hyperref}
\hypersetup{colorlinks=true,
						linkcolor=blue,
						anchorcolor=blue,
						citecolor=blue,
						filecolor=blue,
						menucolor=blue,
						urlcolor=blue}
\usepackage{color}
\usepackage[caption=false]{subfig}
\newcommand{\AC}{$\text{A}_3\text{C}_{60}$\xspace}
\newcommand{\KC}{$\text{K}_3\text{C}_{60}$\xspace}
\newcommand{\RbC}{$\text{Rb}_3\text{C}_{60}$\xspace}
\newcommand{\eN}{\ensuremath{e_{\mathrm{N}}}\xspace}
\newcommand{\Bmax}{\ensuremath{B_{\mathrm{max}}}\xspace}
\newcommand{\N}{\ensuremath{\mathcal{N}}\xspace}

\begin{document}

\title{\texorpdfstring{Superconducting fluctuations observed far above $\text{T}_\text{c}$\\ in the isotropic superconductor \KC}%
                    {Superconducting fluctuations observed far above Tc in the isotropic superconductor K3C60}
}


\author{Gregor Jotzu}
\email{gregor.jotzu@mpsd.mpg.de}
\author{Guido Meier}
\author{Alice Cantaluppi}
\affiliation{Max Planck Institute for the Structure and Dynamics of Matter, Hamburg, Germany}
\author{Andrea Cavalleri}
\affiliation{Max Planck Institute for the Structure and Dynamics of Matter, Hamburg, Germany}
\affiliation{Department of Physics, Clarendon Laboratory, University of Oxford, Oxford, UK}
\author{Daniele Pontiroli}
\author{Mauro Ricc\`{o}}
\affiliation{Dipartimento di Scienze Matematiche, Fisiche e  
Informatiche, Universit\`{a} degli Studi di Parma, Parma, Italy}
\author{Arzhang Ardavan}
\email{arzhang.ardavan@physics.ox.ac.uk}
\author{Moon-Sun Nam}
\email{moon-sun.nam@physics.ox.ac.uk}
\affiliation{Department of Physics, Clarendon Laboratory, University of Oxford, Oxford, UK}

\begin{abstract}
    
{Alkali-doped fullerides are strongly correlated organic superconductors that exhibit high transition temperatures, exceptionally large critical magnetic fields and a number of other unusual properties. 
The proximity to a Mott insulating phase is thought to be a crucial ingredient of the underlying physics, and may also affect precursors of superconductivity in the normal state above ${T_c}$. We report on the observation of a sizeable magneto-thermoelectric (Nernst) effect in the normal state of \KC, which displays the characteristics of superconducting fluctuations. 
The anomalous Nernst effect emerges from an ordinary quasiparticle background below a temperature of 80K, far above ${T_c = 20\text{K}}$.
At the lowest fields and close to ${T_c}$, the scaling of the effect is captured by a model based on Gaussian fluctuations.
The temperature up to which we observe fluctuations is exceptionally high for a three-dimensional isotropic system, where fluctuation effects are usually suppressed.
} 
\end{abstract}

\maketitle


Recent work suggests that the exceptional properties of alkali-doped fulleride superconductors, \AC,  result from an unusual cooperation between electron-phonon and electron-electron coupling\;\cite{nomura2016,capone2002,chakravarty1991}.
The former is primarily governed by a dynamical Jahn-Teller distortion of the $\text{C}_{60}$ molecules leading to an inverted Hund's coupling between electrons\;\cite{[{Phonon effects may not necessarily be required for an inverted Hund's coupling, see }][{}]jiang2016}, whilst the latter contributes to a suppression of the effective bandwidth. 
With increasing lattice spacing, superconductivity in \AC acquires a ``dome-like'' $T_c$, eventually evolving into a Mott insulator with an antiferromagnetic ground state\;\cite{takabayashi2009,zadik2015}.
Unlike other high-temperature superconductors however, \AC features no anisotropy, and displays the characteristics of an s-wave superconductor.
 
\AC also seems to follow the Uemura relation\;\cite{uemura1994,uemura1989}, with a transition temperature $T_c$ proportional to the superfluid density, suggesting that the loss of long-range phase coherence may be responsible for the disappearance of superconductivity at $T_c$. Yet, due to the large spread of experimentally determined superfluid densities in  \AC\;\cite{gunnarsson2004a,kadish2000}, some uncertainty remains on this assignment. 
\begin{figure}
	\includegraphics[width=1\columnwidth]{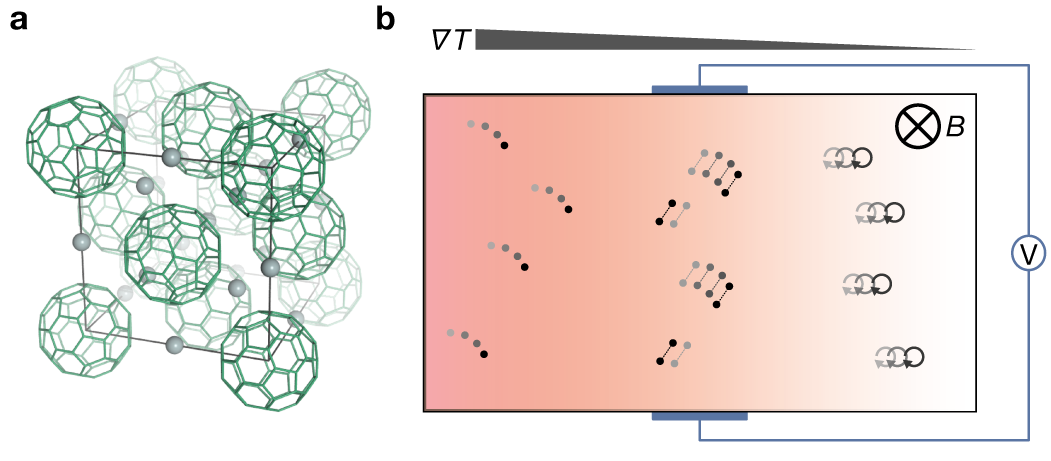}
	\caption{
	\textbf{Probing the Nernst effect in \KC.} 
	\textbf{a,}~The f.c.c. lattice structure of \KC, with potassium atoms shown in grey and carbon buckyballs in green. 
	\textbf{b,}~Schematic of the measurement configuration. A temperature gradient $\nabla T$ is applied orthogonal to the external magnetic field, $B$. The voltage, $V$, is then measured orthogonal to both. Cartoons of various possible contributions to the Nernst signal, such as quasiparticles (left), short-lived Cooper pairs (centre) and mobile vortices (right), are included. 
	}
	\label{fig:1}
\end{figure} 

In \KC (see Fig.~\ref{fig:1}a), observables such as the specific heat and the pressure dependence of $T_c$ suggest that the material may be well described by weak-coupling BCS theory\;\cite{gunnarsson2004a,kadish2000}, but discrepancies in the size and temperature dependence of the superconducting gap remain\;\cite{lieber1994a}. Very recent measurements in few-layer thin films of \KC have reported the appearance of a pseudogap up to about twice $T_c$\;\cite{ren2020}. 

Finally, upon illumination with mid-infrared laser pulses, optical properties compatible with superconductivity have been observed in \KC at temperatures that exceed $T_c$ by an order of magnitude\;\cite{mitrano2016,cantaluppi2018,budden2021,buzzi2021}, further underscoring a highly unusual normal state. 
One of the proposed mechanisms for these phenomena suggests that the effect of the light field consists in synchronizing pre-existing, but phase-incoherent, Cooper pairs\;\cite{budden2021,uemura2019}, inspired by the empirical correlation between materials where light-induced high temperature superconductivity can be induced and an anomalous  Nernst effect above $T_c$ exists\;\cite{uemura2019,nam2007,buzzi2020}.
Recent experiments have also provided suggestive magnetic anomalies when \RbC is made to interact with electromagnetic vacuum modes in an optical cavity\;\cite{thomas2019}. 

The Nernst effect describes the appearance of an electric field, $E_y =-\eN \partial_x T$, transverse to an applied temperature gradient, $\partial_x T$, and to a magnetic field $B_z$ pointing along the third spatial direction. The Nernst signal \eN is related to the conductivity and thermoelectric tensors, $\bm{\sigma}$ and $\bm{\alpha}$ via 
\begin{equation}
e_{\mathrm{N}}= \frac{\alpha_{xy} \sigma_{xx} - \alpha_{xx} \sigma_{xy}}{\sigma_{xx}^2+\sigma_{xy}^2} \approx  \frac{\alpha_{xy}}{\sigma_{xx}} -S \mu_{\mathrm{H}} B_z
\label{eq:eN}
\end{equation}
for an isotropic system. $\mu_\mathrm{H}=\sigma_{xy}/\sigma_{xx}B_z$ denotes the Hall mobility and $S=\alpha_{xx} /\sigma_{xx}$ the Seebeck coefficient. The approximate equality holds for small Hall angles, and is an excellent approximation for the parameters used in this work.

For a metal, the effect can be seen as a combination of a flow of charges carrying entropy along a temperature gradient, the Seebeck effect, and the deflection of moving charges in the presence of a magnetic field, the Hall effect (see Fig.~\ref{fig:1}b). However, with exact particle-hole symmetry, the two terms in Eq.~\ref{eq:eN} would cancel exactly\;\cite{behnia2009}. The overall sign and amplitude of the effect will depend on the details of the quasiparticle band structure, which the Nernst signal is therefore very sensitive to. In the free-electron approximation, this signal is linear in temperature.

In a superconductor, a different contribution to the Nernst effect arises from the movement of superconducting vortices. When these mobile vortices carry entropy along the applied temperature gradient (see Fig.~\ref{fig:1}b), they also carry magnetic flux, which induces a voltage in the transverse direction\;\cite{behnia2016,[{An additional contribution from the Magnus force is also present, see }][{}]krasnov1997}. This effect is much larger than its metallic counterpart, and shows a highly non-linear dependence on $B_z$, perhaps owing to a competition between vortex density and vortex mobility. 

If, at $T_c$, superconducting long-range order breaks down because of fluctuations of the phase of the order parameter, while a finite Cooper pair amplitude remains, this vortex Nernst effect would be expected to survive at temperatures above $T_c$. However, even for a transition driven by a thermal breakdown of Cooper pairing, the thermal diffusion of short-lived Cooper pairs may also contribute to the Nernst signal above $T_c$\;\cite{ussishkin2002,pourret2006,behnia2016}.
 
In general, the presence of such precursors of superconductivity is expected to be suppressed as the dimensionality of the system increases. Indeed, Nernst signals that could be related to superconducting precursors have been reported in layered and thin-film samples\;\cite{ong2004,pourret2006,nam2007,cyr-choiniere2018,behnia2016}, but we are not aware of any previous observations in an isotropic three-dimensional system. Interestingly, \KC and \RbC were the first fully three-dimensional materials where slight deviations of the normal state conductivity in the immediate vicinity of $T_c$ could be attributed to paraconductivity\;\cite{xiang1993}.

\begin{figure}
	\includegraphics[width=1\columnwidth]{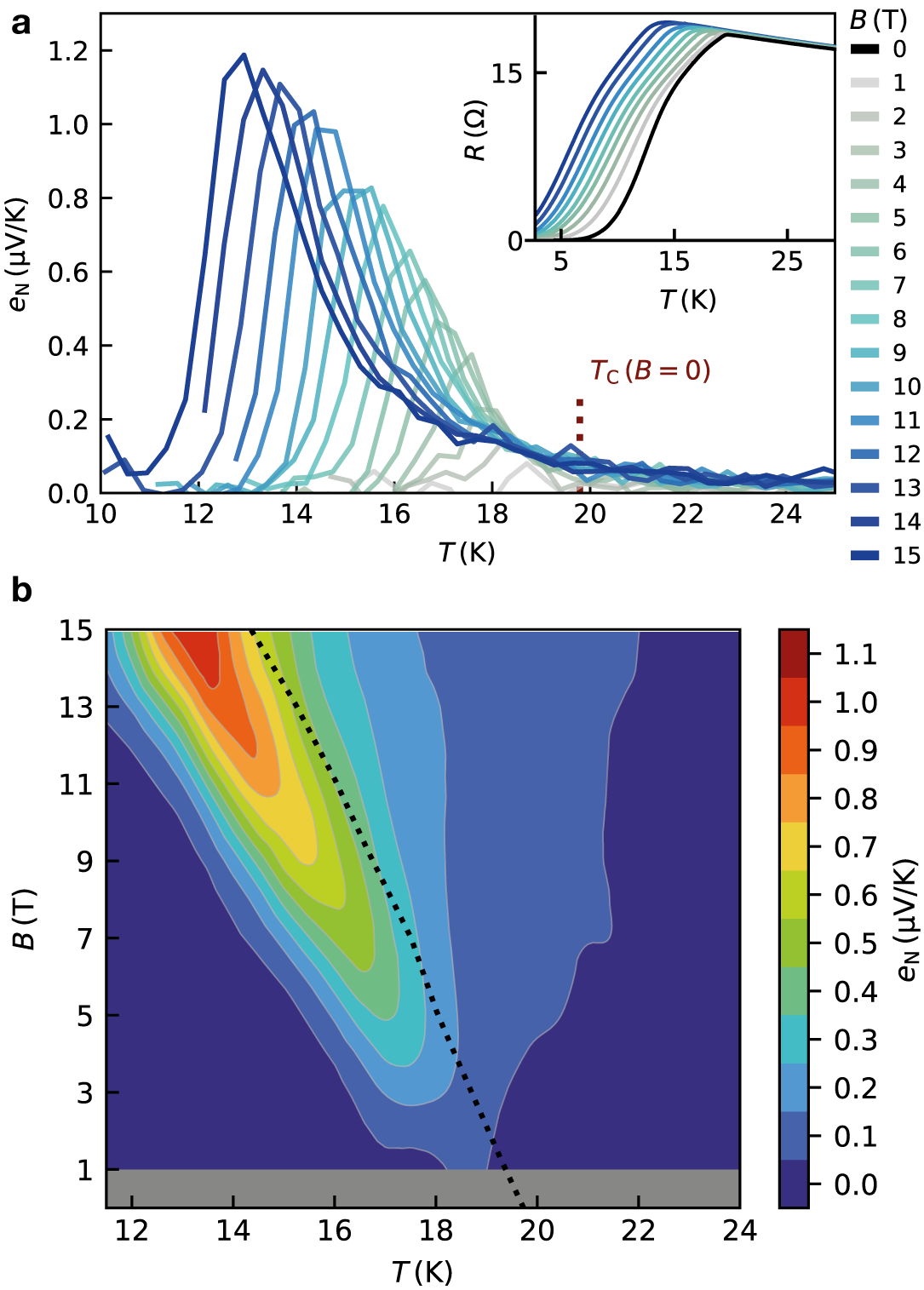}
    \pagestyle{empty} 
	\caption{
	\textbf{Nernst signal in the superconducting regime.} 
	\textbf{a,}~Measured transverse thermoelectric effect, \eN, in the presence of magnetic fields ranging from 1T (grey line) to 15T (blue line), in steps of 1T. The red dashed line shows the critical temperature in zero field. Inset: Sample resistance near the critical temperature, 0T (black) to 15T (blue) range. 
	\textbf{b,}~Contour plot of \eN as a function of temperature and magnetic field. The black dotted line shows the resistive $T_c(B)$.  A Gaussian convolution was used for smoothing, and in the low-T, low-B region the data was sparse and was interpolated. See Fig.~\ref{fig:S3}b for raw data and the field-normalized $\eN/B$.
	}
	\label{fig:2}
\end{figure}

In the experiments reported in this paper, air-sensitive \KC powders were compressed into pellets and incorporated into a circuit board printed on an FR4 substrate, which features low thermal conductivity. Embedded heaters and temperature sensors, as well as indium-coated contacts with contact resistances below 1 Ohm were used to optimize these measurements (see Appendix and Fig.~\ref{fig:S1}). Four-probe resistance measurements (see Appendix) were used to identify the superconducting transition, as shown in the inset of Fig.~\ref{fig:2}a. In zero field, we found $T_c(B=0) = 19.8$K, in good agreement with previous reports and with magnetic measurements on the same batch of samples\;\cite{gunnarsson2004a,kadish2000,cantaluppi2018}. A Werthamer-Helfand-Hohenberg theory\;\cite{werthamer1966} was used to extrapolate the zero-temperature upper critical field, $\mu_0 H_{c2}(0)$, 
from the field dependence of the resistive transition via: $\mu_0 H_{c2}(0) = 0.69 \mu_0 T_c \left. \frac{\partial H_{c2}}{\partial T} \right|_{T_c}$, where $\mu_0$ denotes the vacuum permeability. This yielded a value of about 39T, corresponding to a zero-temperature coherence length of $\xi_0 = \sqrt{\Phi_0/2\pi\mu_0 H_{c2}(0)} =2.9$nm (here $\Phi_0$ is the magnetic flux quantum). This lies within the range of values extrapolated from other experiments\;\cite{gunnarsson2004a,kadish2000}, and closely matches the value recently determined using pulsed fields in a range exceeding $H_{c2}$ \;\cite{kasahara2017}.

As expected, no Nernst signal (\eN) was observed for temperatures far below $T_c$, likely due to freezing of vortex motion. 
For higher temperatures, entering the temperature range where vortices become mobile, the signal was seen to increase (see Fig.~\ref{fig:2}a). 
For a given magnetic field, the number of vortices in the system remains nearly constant (as all fields in our measurements were much larger than the lower critical field), but their mobility increases rapidly. 
Near $T_c$, the Nernst signal reduced again. 
A detailed quantitative understanding of this well-known phenomenon is still lacking -- both the increasing vortex-vortex interactions that appear as vortex length scales increase, as well as a change in the entropy per vortex are relevant\;\cite{behnia2016,podolsky2007}. 
Although the amplitude of \eN did not saturate at the largest applied field, the value it approaches agrees with an upper bound that was found to apply to various superconductors spanning several orders of magnitude in critical fields and temperatures\;\cite{rischau2021}. 

\begin{figure}
	\includegraphics[width=1\columnwidth]{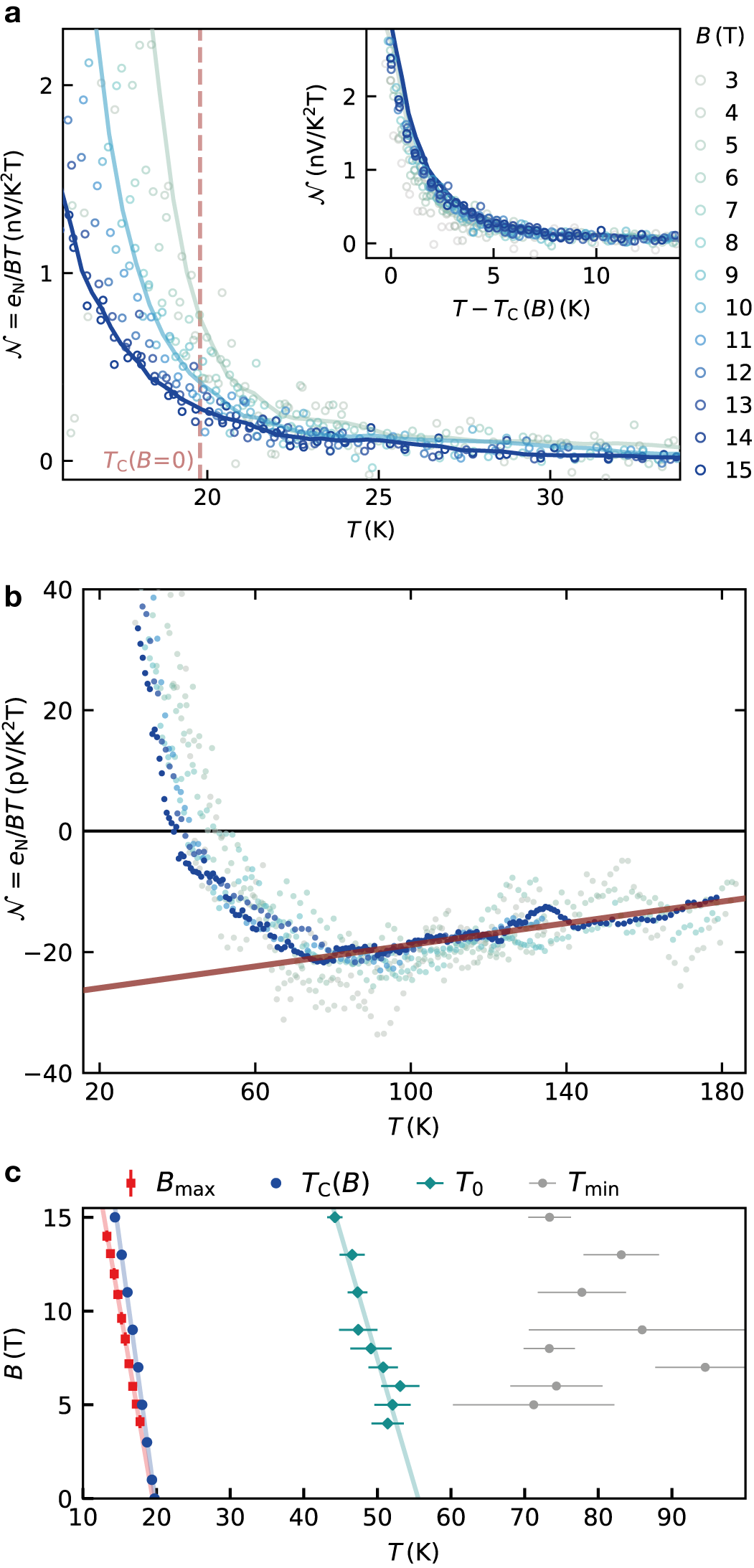}
	\caption{
	\textbf{Anomalous Nernst effect above $T_c$.}
	\textbf{a,}~Nernst coefficient \N  close to $T_c$ (red dashed line). Open symbols show the raw data, lines are 10-point boxcar averages for 5T, 10T and 15T (grey to blue). Inset: Same data re-scaled with the field-dependent $T_c(B)$ .
	\textbf{b,}~\N~at higher temperatures, after 10-point boxcar averaging (Note the reduced $y$-scale). The red curve corresponds to the quasiparticle contribution $-2.7\mu_\mathrm{H}S/T$.
	\textbf{c,}~Characteristic quantities extracted from the Nernst signal: \Bmax denotes the fitted peak of $\eN(B)$ (see Fig.~\ref{fig:2}b), $T_0$ and $T_\mathrm{min}$ are the zero-crossing and minimum of \N, respectively. $T_c$ denotes the resistive superconducting transition temperature. All lines are linear fits. Error bars indicate fit uncertainties.
	}
	\label{fig:3}
\end{figure}

The Nernst signal as a function of magnetic field (see  Fig.~\ref{fig:S3}a) peaked at a field \Bmax, that reduces on approaching the critical temperature. 
This reduction fits very well to a linear function with a slope of $-2.27(3)$\,T/K, see Fig.~\ref{fig:3}c. It displays a zero-field intersect at $T= 19.5(1)$\,K, close to the resistive critical temperature $T_c(B=0)=19.8$K. 
\Bmax has previously been shown to bear some similarity to a softening mode\;\cite{tafti2014,behnia2016}, and its vanishing value when approaching $T_c(0)$ suggests that the Nernst effect changes in nature when crossing $T_c$.


Above $T_c$, precursors of superconductivity may contribute to the Nernst signal, but quasiparticles can also play a role. The latter contribution is expected to scale linearly with magnetic field and temperature (at constant volume) for a simple metallic state.\;\cite{behnia2009} We therefore used the temperature- and magnetic-field-normalized Nernst coefficient, $\N = \eN/BT$, in order to detect anomalous behaviour. In Fig.~\ref{fig:3}a, \N is shown to evolve smoothly across $T_c$, remaining positive and retaining a strong field dependence, as expected for a signal that is primarily caused by superconducting fluctuations. The inset in Fig.~\ref{fig:3}a shows that \N displays near-universal behaviour upon approaching the field-dependent critical temperature, $T_c(B)$, (see also Fig.~\ref{fig:S3}d).

At higher temperatures, we observed two characteristic features in the data (see Fig.~\ref{fig:3}b): First, at $T_0(B) \approx 50$K, the signal changed from positive to negative. The temperature of the zero-crossing, $T_0(B)$, shows a similar field dependence as $T_c(B)$, (see Fig.~\ref{fig:3}c).  Second, a minimum in \N appears at $T_\mathrm{min}(B) \approx 80$K, above which \N shows a linear and positive slope. 

In order to disentangle the superconducting contribution to \N we first focus on the high-temperature limit, where the quasiparticle contribution should become dominant. We can compare the behaviour of \N in this regime to the expected signal in a single band free-electron model of a metal. There, $|\N|\sim\mu_\mathrm{H} S/T$, with a pre-factor of order unity\;\cite{behnia2009}. This relationship implies that the Nernst effect is proportional to the ratio of the mobility and the Fermi energy in the metal, which was found to hold in a variety of materials, with \N ranging from 1mV/$\mathrm{K}^2$T down to 1nV/$\mathrm{K}^2$T\;\cite{behnia2016}. We use previously determined values for the $T$-linear $S$ and $\mu_\mathrm{H}$\;\cite{inabe1992,lu1995}, where the latter shows a linear dependence on temperature that has been shown to scale with the expansion of the lattice. As shown in Fig.~\ref{fig:3}b, we find excellent agreement with our data above $T_\mathrm{min}$, using a quasiparticle contribution of $-2.7\mu_\mathrm{H}S/T$, and no significant field-dependence of \N. Given the complex band structure of \KC, the agreement with such a simple scaling is remarkable. It extends its range of validity to values of \N which are two orders of magnitude smaller than those reported so far\cite{behnia2016}.

The rapid change in the slope of \N signals a significant change in the electronic properties of the material. Such a change could for example be caused by the appearance of charge-density wave order\;\cite{bel2003}, and in a number of cuprate superconductors a very similar feature was found to coincide closely with the pseudogap temperature $T^*$\;\cite{cyr-choiniere2018}. In \KC, a transition to a frozen orientational disorder of the $\mathrm{C}_{60}$ molecules is known to occur, but at a temperature very close to 200K\;\cite{yoshinari1993,goldoni1999}. There are no observations pointing at the appearance of competing orders around 80K, although it is worth noting that a certain deviation from linearity has been observed in the Seebeck effect, which has been attributed to electron-phonon coupling or precursors of superconductivity\;\cite{inabe1992,sugihara1993,morelli1994}. 

An effect related to the superconducting ground state seems much more likely, given that the magnetic field dependence of \N tracks $T_c(B)$. In the following we will consider two scenarios through which precursors of superconductivity could cause such an upturn in \N: a vortex-based Nernst signal surviving in a phase-fluctuating regime above $T_c$, or a signal caused by short-lived Cooper pairs. 

\begin{figure}
	\includegraphics[width=1\columnwidth]{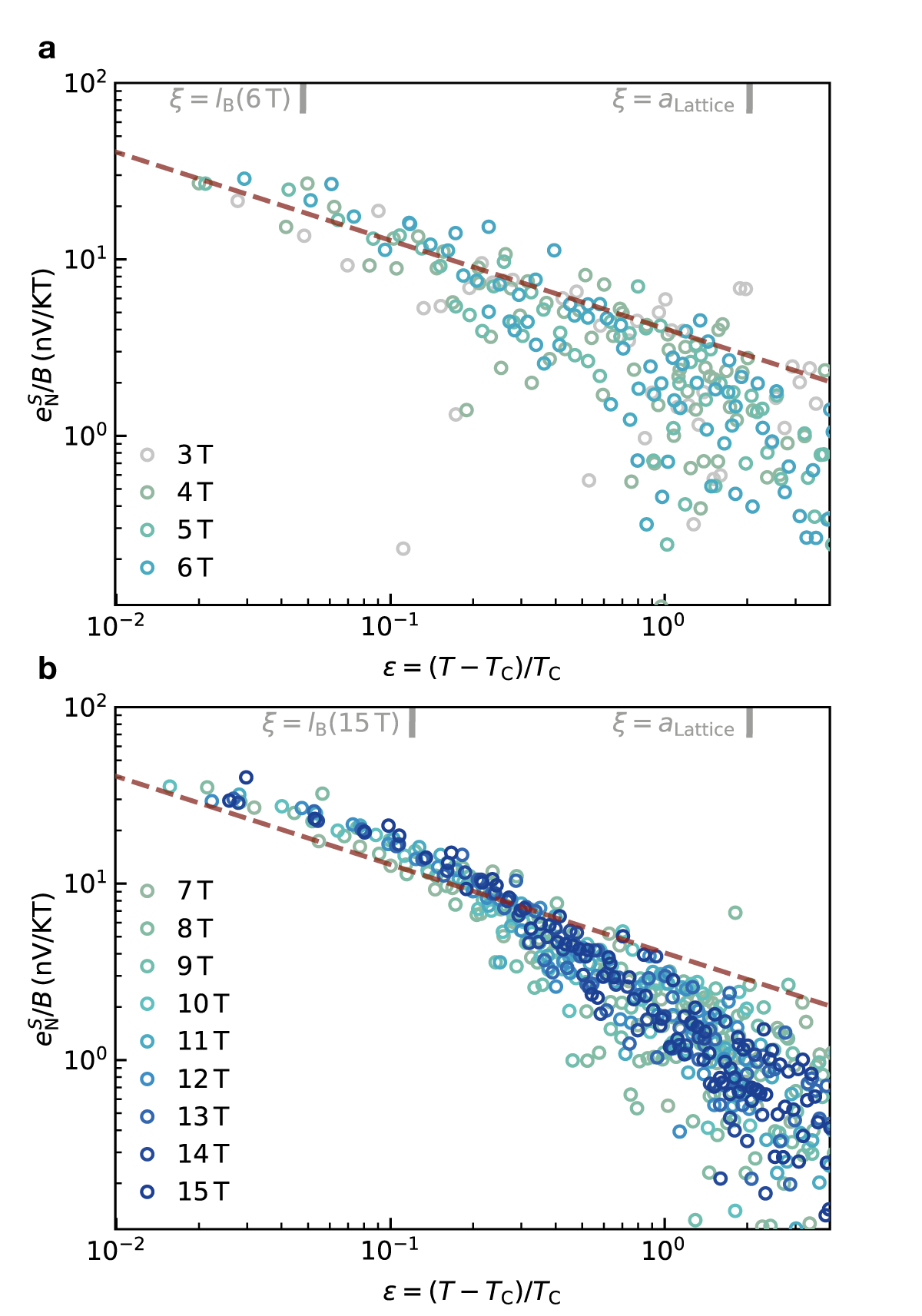}
	\caption{
	\textbf{Scaling of the superconducting Nernst coefficient $\eN^S/B$ above $T_c$.} 
	\textbf{a,}~Quasiparticle-subtracted contribution to the Nernst coefficient, $\eN^S/B=\eN/B + 2.7\mu_\mathrm{H}S$, for magnetic fields between 3T and 6T. Color scale as in Fig.~\ref{fig:1}. The data are shown as a function of the distance to the field-dependent critical temperature at zero field.
	\textbf{b,}~Data for higher fields, up to 15T. The dashed red line shows the expected low-field value from a model based on Gaussian fluctuations, with a constant conductivity of 6$(\mathrm{m}\Omega\mathrm{cm})^{-1}$ (see text). The top axes indicate where the coherence length, $\xi$, becomes equal to the lattice spacing, $a_\mathrm{Lattice}$, or to the magnetic length $l_B$ at the highest field shown in the respective panel.
	}
	\label{fig:4}
\end{figure} 

For the latter scenario, a theory based on Gaussian fluctuations in a Ginzburg-Landau model\;\cite{ussishkin2002} predicts a superconducting contribution, $e_{\mathrm{N}}^{\mathrm{SCG}}$, to the Nernst signal given by:
\begin{equation}
\label{eq:Gauss}
\frac{e_{\mathrm{N}}^{\mathrm{SCG}}}{B_z} = \frac{\alpha_{xy}^{\mathrm{SCG}}}{\sigma_{xx}B_z} = \frac{k_\mathrm{B} e^2}{12\pi\hbar^2}\frac{\xi}{\sigma_{xx}}  
\end{equation}
\quad\quad\text{with}\quad
\begin{equation}
\xi = \frac{\xi_0}{\sqrt{\epsilon}} = \frac{\xi_0}{\sqrt{(T-T_c)/T_c}}
\end{equation}
for a three-dimensional system, where $e$ is the electron charge and $\hbar$ the reduced Planck constant. The two-dimensional version of this theory is in good agreement with measurements on conventional\;\cite{pourret2006,behnia2016} and some unconventional\;\cite{chang2012,tafti2014} superconductors. Its prediction of a field-independent Nernst coefficient applies to the low-field regime, where the coherence length $\xi$ is short compared to the magnetic length $l_B = \sqrt{\hbar/eB}$, and $T_c(B)\approx T_c(0)$. As it is a continuum theory, it may also become invalid once $\xi$ becomes as short as the lattice spacing, and generally speaking Ginzburg-Landau theory is only applicable in the vicinity of $T_c$.

In Fig.~\ref{fig:4}a, we compare our measurements for fields up to 6T to this theory by subtracting the fitted quasiparticle contribution determined above (red line in Fig.~\ref{fig:3}b) from the measured Nernst signal. See Fig.~\ref{fig:S2} for other possible subtraction schemes. We find that the data is overall well described by the simple $1/\sqrt{\epsilon}$ scaling predicted by theory if we use a conductivity of 6$(\mathrm{m}\Omega\mathrm{cm})^{-1}$. Although we expect our measurement to be sensitive to the intrinsic conductivity, rather than grain boundary effects\;\cite{kang2017}, this value is still about three times larger than the conductivity of high-purity single crystals.\;\cite{gunnarsson2004a} Interestingly however, a very similar discrepancy was found in measurements of the  paraconductivity\;\cite{xiang1993}, suggesting that the residual conductivity describing the transport properties of superconducting fluctuations may not be identical to the one extracted from direct measurements.

At higher temperatures, beyond the expected regime of validity of this model, deviations from the simple scaling occur, which might be captured by perturbative expansions\;\cite{michaeli2009,serbyn2009}. Interestingly, at higher fields (see Fig.~\ref{fig:4}b), the behaviour does not follow a single power law. This suggests that in high field, the resistive $T_c(B)$ deviates from the thermodynamical critical point, which is a characteristic of unconventional superconductors \cite{blatter1994}. 

In order to gauge the plausibility of a phase-fluctuating scenario, we use the framework proposed by Emery and Kivelson\;\cite{emery1995} to estimate the temperature $T_\theta$ at which global phase coherence in the superconductor would be destroyed by thermal fluctuations -- even if pairing were to survive up to a higher ``mean field temperature'' $T_\mathrm{MF}$. Taking the most recent (and largest) value for the penetration depth in \KC, $\lambda=890$nm\;\cite{buntar1996a}, we find a temperature $T_\theta$ as low as 80K. This is only 4 times larger than $T_c$, whereas for conventional superconductors $ T_\theta/T_\mathrm{MF}$ can be on the order of $10^5$. Additionally, by taking into account some degree of quantum fluctuations (as expected given the relative proximity of a Mott-insulating state), it would be possible that the superconducting transition is somewhat suppressed below $T_\mathrm{MF}$. 

We are not aware of a quantitative prediction of the vortex Nernst signal above $T_c$ in a three-dimensional system. Results in two dimensions\;\cite{podolsky2007} suggest an important role of the lattice geometry, and can therefore not simply be extrapolated to our system. As the difference between mean-field models and fully quantum-mechanical descriptions becomes less pronounced in higher dimensions, small deviations at high temperatures and magnetic field may play a key role in distinguishing different scenarios. 
Our data therefore also provide an important benchmark for a  theoretical framework describing the appearance of light-induced superconductivity in \KC based on the synchronization of stable but globally phase-incoherent Cooper pairs, which would also have to correctly describe the initial static state.

It would be highly interesting to extend our work to \RbC and especially $\text{Rb}_x\text{Cs}_{3-x}\text{C}_{60}$, where quantum phase fluctuations caused by the proximity of the Mott-insulating state will be enhanced. For the latter family, a suppression in $T_c$ upon approaching the quantum phase transition has been observed\;\cite{takabayashi2009,zadik2015}, but could not be reproduced in an otherwise quantitatively successful theoretical model\;\cite{nomura2016}. Studying the Nernst effect in this regime, which should be possible using the experimental framework presented here, would provide new insights concerning the nature of the superconducting transition in the fullerides, and of phase-incoherent superconductivity in general. 

\textbf{Acknowledgements} 
We thank Michele Buzzi, Dante Kennes, Daniel Podolsky and Dharmalingam Prabhakaran for insightful discussions, and Boris Fiedler for superb technical assistance.

\vspace{0.5cm}


\makeatletter
\setcounter{section}{0}
\setcounter{subsection}{0}
\setcounter{figure}{0}
\setcounter{table}{0}
\setcounter{NAT@ctr}{0}

\renewcommand{\thefigure}{S\@arabic\c@figure}

\appendix
\section*{Appendix} 
\section{Sample preparation}
The \KC powder used in this work was prepared and characterized as previously reported in Refs. \cite{mitrano2016,cantaluppi2018,budden2021}. In brief, finely ground C$_{60}$ powder and metallic potassium were placed in a vessel inside a Pyrex vial in stoichiometric amounts, evacuated to $10^{-6}$mbar, and sealed. The two materials were heated at 523K for 72h and then at 623K for 28h, and kept separated to ensure that the $C_{60}$ was only exposed to clean potassium vapour. After regrinding and pelletizing in an inert Ar atmosphere, the sample was annealed at 623K for 5 days. Powder X-ray diffraction measurements confirmed the purity of \KC and indicated a grain size between 100nm and 400nm. Magnetic susceptibility measurements yielded a $T_c$ of 19.8K\cite{mitrano2016}.
For the Nernst effect and 4-point resistivity measurements the sample was handled inside an Ar glove box with $<$0.2ppm $\text{O}_2$ and $\text{H}_2$O. It was placed inside an FR4-frame, which had been glued to the circuit board described below using thermally and electrically insulating, minimally-outgassing glue (Epo-TeK 301-2FL-T), see Fig.~\ref{fig:S1}. The powder was then compressed with an FR4 piston, and sealed with the same glue. The resistance of the sample was monitored to ensure that no contamination occurred during the sealing process and the subsequent transfer to the cryostat.
\section{Nernst effect measurement setup}

\begin{figure}
	\includegraphics[width=1\columnwidth]{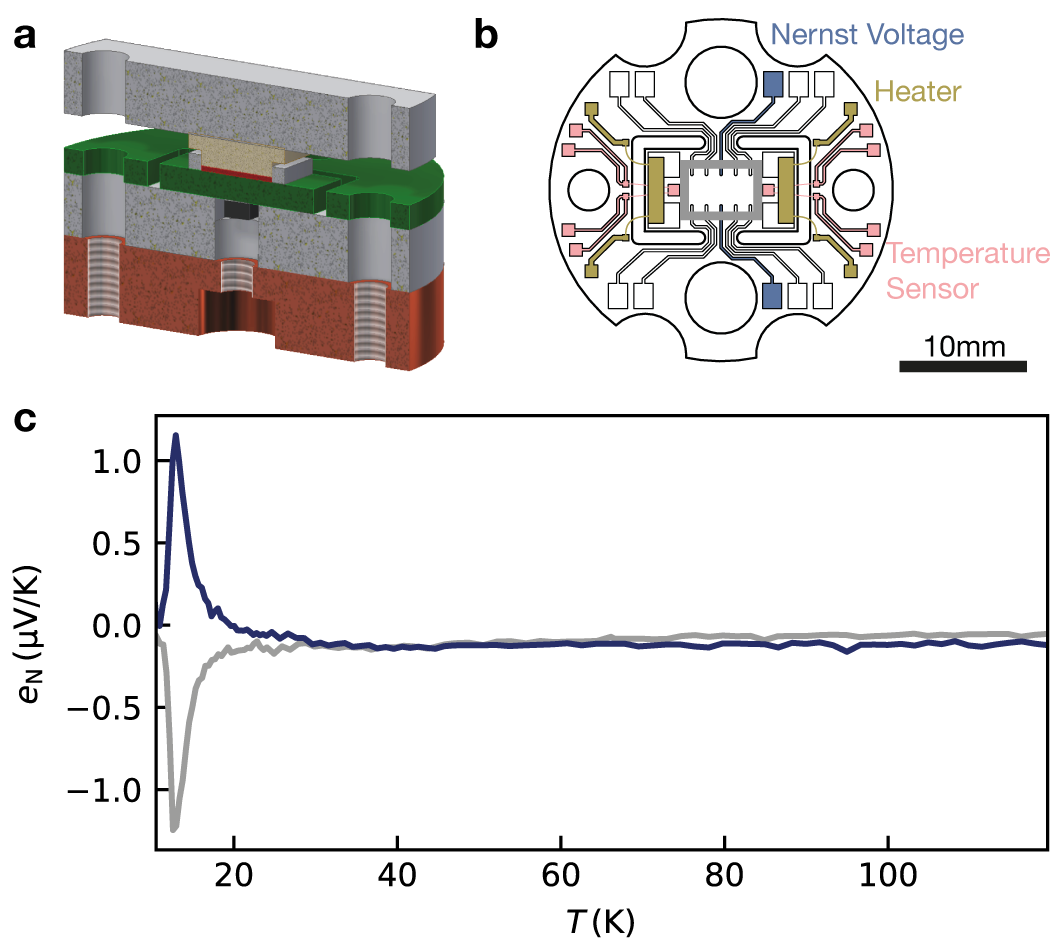}
	\caption{
	\textbf{Experimental setup.} 
	\textbf{a,}~The sample (red) is pressed on top of a circuit board (green) using a PMMA piston (beige). The sample space is encapsulated using non-conductive epoxy glue. The base temperature is measured with a sensor (black) pressed against the bottom of the circuit board. The sample and circuit board are mounted on the cold finger of a cryostat (copper) using titanium screws (not shown), using spacers made of PMMA (grey). 
	\textbf{b,}~Schematic of the circuit board. The sample compartment is indicated by a grey rectangle. A temperature sensor is placed on each side of the sample in a milled pocket, encapsulated by thermally-conductive epoxy glue. Resistive heaters are placed on each side. The Nernst signal is measured with the indicated indium-coated contacts in the centre of the sample compartment. The other contacts are used for four-point resistance measurements.
	\textbf{c,}~Transverse voltage at +15T (blue) and -15T (grey). For the data shown in the main text, the Nernst signal was evaluated as half of the difference between signals measured at opposite fields.
	}
	\label{fig:S1}
\end{figure}
We used a printed copper circuit board on an FR4 substrate (which features a low thermal conductivity, $\sim 0.1$W/Km at 10K\;\cite{woodcraft2009}), as shown in Fig.~\ref{fig:S1}b. Cernox temperature senors were embedded in thermally conductive glue  (Stycast 2850FT) in milled pockets on each side of the sample. They were used to monitor the temperature gradient across the sample, which was induced using a resistive heater. An additional Cernox sensor was attached to the bottom of the circuit board to monitor the base temperature. The circuit board was mounted to the cold finger of a cryostat using non-magnetic (titanium) screws and spring washers, and PMMA spacers were used for additional thermal insulation. In the sample compartment, the copper contacts were coated with indium, yielding contact resistances below 1 Ohm. The transverse voltage was measured whilst slowly cooling the sample. Data for opposite magnetic fields (see Fig.~\ref{fig:S1}c), was then subtracted to compute the Nernst signal.
\section{Resistance measurements}
The resistance of the sample was determined using a low-frequency lock-in measurement in a linear 4-contact configuration, with contacts as shown in Fig.~\ref{fig:S1}b. Above $T_c$, the sample showed an increase in resistance upon cooling, as previously observed in granular \KC samples\;\cite{gunnarsson2004a,kadish2000}. In order to determine $T_c$, the point at which the resistance changes slope was used, which yielded a zero-field $T_c$ consistent with magnetic susceptibility measurements on the same batch of sample. We verified that the width of the transition is not sensitive to reducing the probe current below the value of 2$\mathrm{\mu}$A which we used.
\begin{figure}
	\includegraphics[width=1\columnwidth]{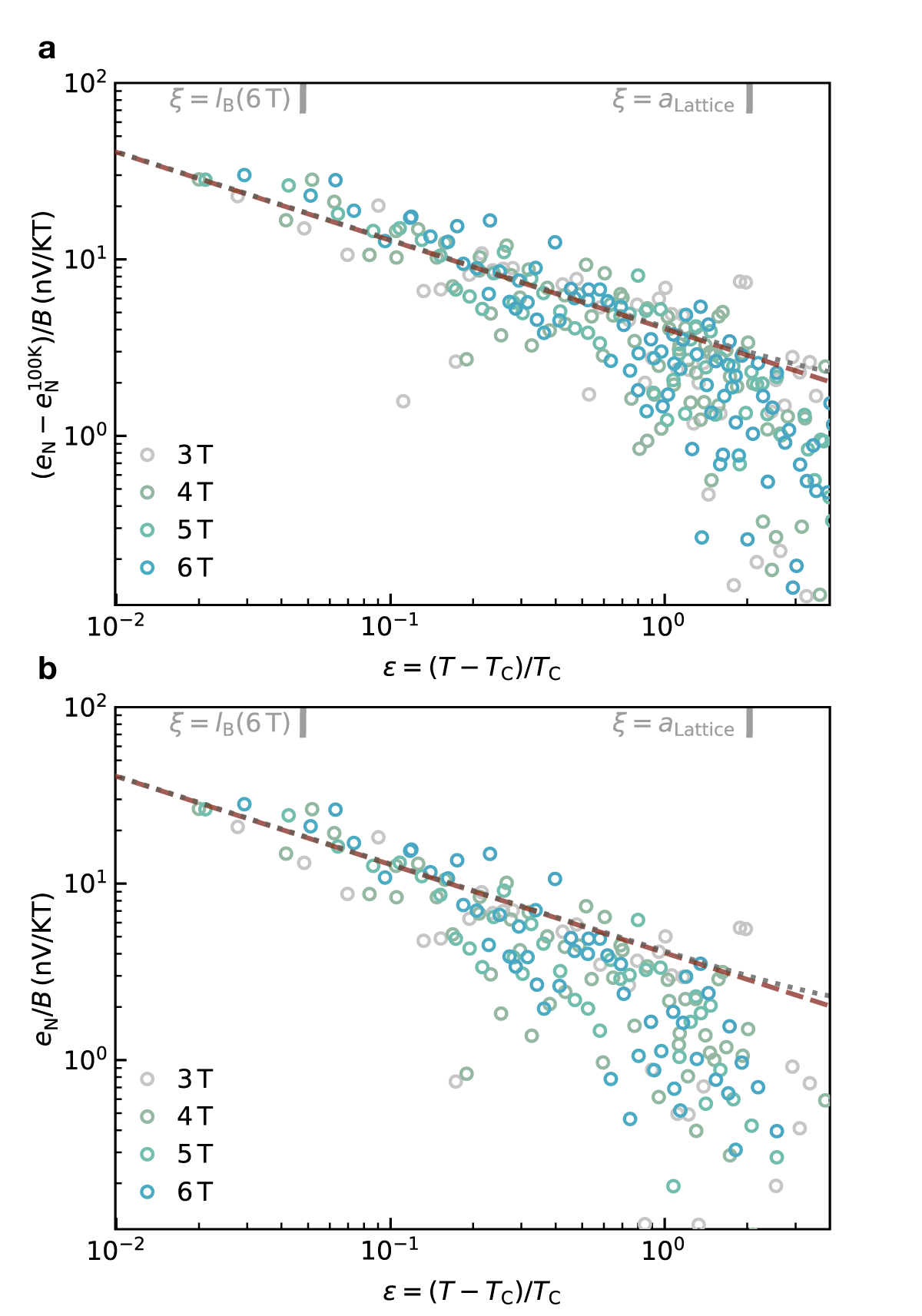}
	\caption{
	\textbf{Effect of subtracting the quasiparticle signal on the scaling analysis.} 
	\textbf{a,}~Same as Fig.~\ref{fig:4}a, but subtracting the fixed value of \eN /B at 100K instead of a temperature-dependent function. The grey dotted line shows the theoretical prediction of Eq.~\ref{eq:Gauss}, but using a quadratic temperature dependence for the conductivity. 
	\textbf{b,}~The Nernst coefficient \eN /B without any subtraction of the quasiparticle contribution. Note that this results in some negative values for \eN at higher temperatures, which do not appear in this logarithmic plot.
	}
	\label{fig:S2}
\end{figure}
\section{Subtraction of the quasiparticle signal}
\begin{figure*}
	\includegraphics[width=2\columnwidth]{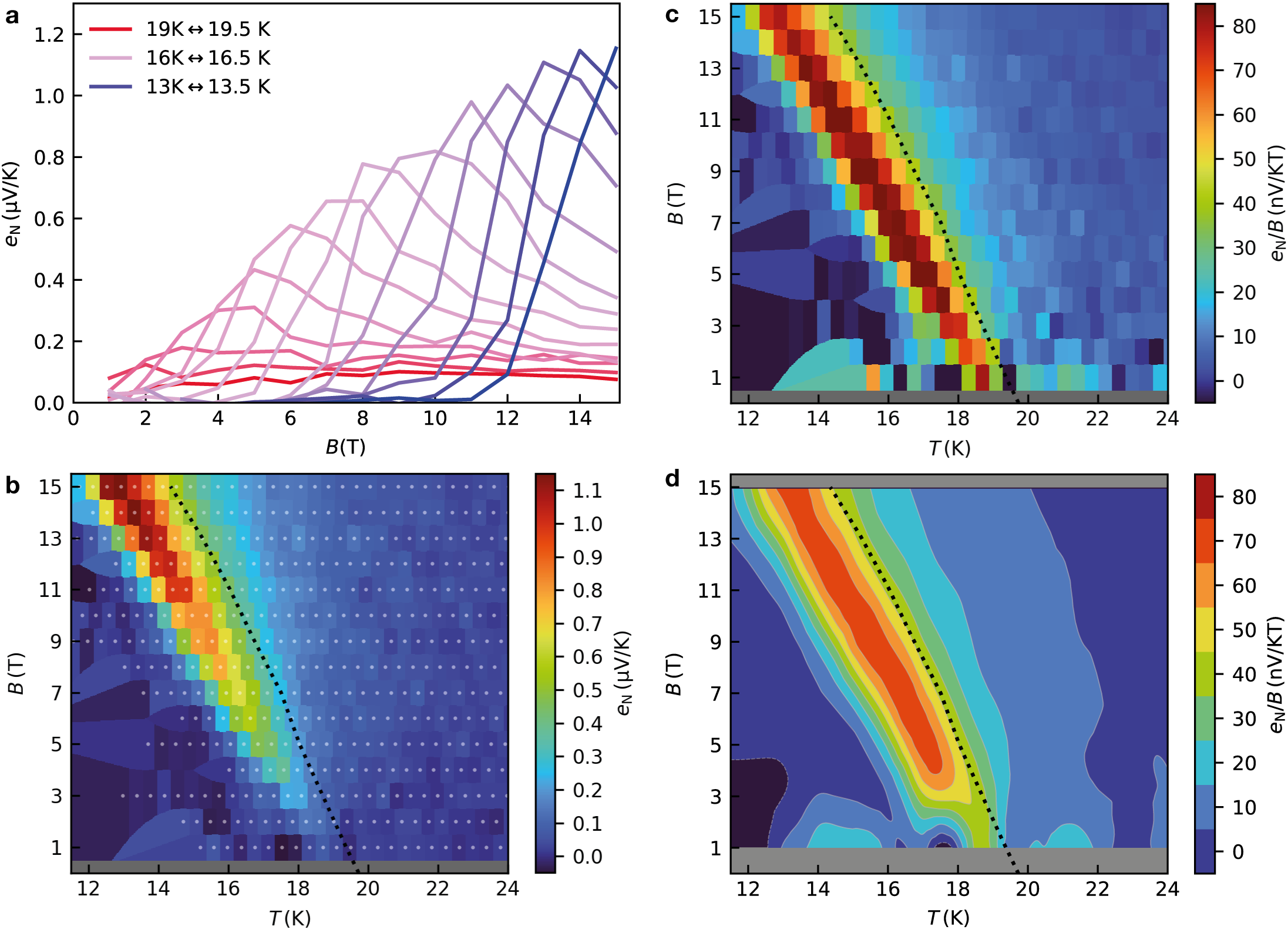}
	\caption{
	\textbf{B-T maps of the Nernst signal.} 
	\textbf{a,}~\eN as a function of magnetic field for different temperatures below $T_c(B)$.
	\textbf{b,}~Raw data of \eN as a function of temperature and magnetic field (see Fig.~\ref{fig:2} for a smoothed contour map). Grey dots indicate the B-T values of each measurements, the data is interpolated to the nearest available point. The black dotted line shows $T_c(B)$.
	\textbf{c,}~Same as (b), but showing the field-normalized value $\eN/B$.
	\textbf{d,}~Smoothed contour plot of (c).
	}
	\label{fig:S3}
\end{figure*}

We verify that the details of how the quasiparticle contribution to \eN is subtracted do not strongly affect the comparison to the Gaussian fluctuation model shown in Fig.~\ref{fig:3}, by comparing different subtraction schemes: In Fig.~\ref{fig:S3}a, instead of subtracting the temperature-dependent quasiparticle function ($-2.7\mu_\mathrm{H}$S), we subtract its fixed value at 100K. Here, the simple theoretical model seems to capture the data even at higher temperatures.
In Fig.~\ref{fig:S3}b, we plot \eN without any quasi-particle subtraction. This leads to a deviation at high temperatures (as expected given that the signal changes sign there), but within the expected range of validity of the theoretical model, it still captures the data well.
We have also verified that using a temperature-dependent value of the conductivity in Eq.~\ref{eq:Gauss} \,(where we have used the quadratic dependence found in\;\cite{klein1992a} as a comparison), has a negligible effect in the relevant range.


%

\end{document}